\documentclass[12pt,a4paper]{article}
\usepackage{float}
\usepackage{amsmath}
\usepackage{amstext}
\usepackage[psamsfonts]{amssymb}
\usepackage[unicode,bookmarks,bookmarksopen,bookmarksopenlevel=0,colorlinks,%
plainpages=false,pdfpagelabels,hypertexnames=false,pageanchor=true,breaklinks=true,%
linkcolor=blue,citecolor=blue,urlcolor=red]{hyperref}
\usepackage[left=2.0cm, right=2.0cm, top=1.0cm, bottom=1.5cm,
includehead]{geometry}
\usepackage{epsfig}
\usepackage{upref}  
\usepackage{indentfirst}
\usepackage[mathcal]{euscript}
\usepackage[cp1251]{inputenc}
\usepackage{version}
\usepackage{theorem}

\newcommand\ben{\begin{enumerate}}
\newcommand\een{\end{enumerate}}
\newcommand\bed{\begin{itemize}}
\newcommand\eed{\end{itemize}}
\newcommand\bei{\begin{description}}
\newcommand\eei{\end{description}}
\newcommand\be{\begin{equation}}
\newcommand\ee{\end{equation}}
\newcommand\beq{\begin{eqnarray}}
\newcommand\eeq{\end{eqnarray}}
\newcommand\beqo{\begin{eqnarray*}}
\newcommand\eeqo{\end{eqnarray*}}

\theoremstyle{plain}\theorembodyfont{\rm}

\newtheorem{Exmpl}{Example}
\newtheorem{Remark}{Remark}

\newcommand\const{\operatorname{const}}

\def\pr{{\partial}}
\def\eps{{\epsilon}}

\def\be{{\bf e}}

\newcommand\pt{\partial}

\begin{document}
\thispagestyle{empty}
\begin{center}
{\large \textbf{  Drift, stabilizing and destabilizing for a Patlak-Keller-Segel system with the short-wavelength external signal }}

\bigskip
{\large \textbf{Andrey Morgulis}\footnote{corresponding author}}
\\
{Southern Mathematical Institute of VSC RAS, Vladikavkaz, Russia\\
 I.I.Vorovich Institute for Mathematic, Mechanics and Computer Science,\\
  Southern Federal University, Rostov-na-Donu, Russia}
\\
\texttt{{Email:morgulisandrey@gmail.com}}

\smallskip
{\large and}

\smallskip
{\large \textbf{Konstantin Ilin}}
\\
{Dept. of Math, The University of York, Heslington, York, UK}
\\
{\texttt{Email:konstantin.ilin@york.ac.uk }}

\bigskip
\begin{minipage}{150mm}
\noindent
{\footnotesize
\begin{center}
\textbf{Abstract.}
\end{center}
This article aims at exploring the short-wavelength stabilization and destabilization of the advection-diffusion systems formulated using the Patlak-Keller-Segel cross-diffusion. We study a model of the taxis partly driven by an external signal.  We address the general short-wavelength signal using the homogenization technique, and then we give a detailed analysis of the signals emitted as the travelling waves.  It turns out that homogenizing produces the drift of species, which is the main translator of the external signal effects, in particular, on the stability issues. We examine the stability of the quasi-equilibria - that is, the simplest short-wavelength patterns fully imposed by the external signal. Comparing the results to the case of switching the signal off allows us to estimate the effect of it. For instance, the effect of the travelling wave turns out to be not single-valued but depending on the wave speed. Namely, there is an independent threshold value such that increasing the amplitude of the wave destabilizes the quasi-equilibria provided that the wave speed is above this value. Otherwise, the same action exerts the opposite effect. It is worth to note that the effect is exponential in the amplitude of the wave in both cases.}

\end{minipage}
\end{center}
\section*{Introduction}
\label{Intro}
\addcontentsline{toc}{section}{Introduction}
\noindent  In this article, we address PDE systems formulated using the Patlak-Keller-Segel (PKS) law.\footnote{The Patlak-Keller-Segel  law claims that the flux of the biological substance pursuing/evading some scalar is everywhere parallel to the density gradient of this scalar.  The ability of a biological substance to such pursuing/evading  is called taxis. The object of the pursuit/ evasion is called the stimulus or signal.}
  Studying such systems mainly aims at modelling the pattern formation in the biological media such as the communities of living species or the biological tissues.  A reader can follow the state of the research starting from review \cite{Ts94} to the recent review \cite{BellBellTao}.


Typically, the PKS system is a reaction-advection-diffusion system into which the PKS law brings the specific nonlinear cross-diffusion. A homogeneous, or, equivalently, translationally invariant PKS system usually can stay in an equilibrium which is homogeneous too, in the sense that the density of every species is constant. Bringing the system out  of such an equilibrium is necessary for occurring  the non-trivial spatiotemporal patterns. An evident reason for getting out the equilibrium is the instability of it to small perturbations. In such an occasion,  the non-trivial patterns can arise from the local bifurcation accompanying the occurrence of the instability.  At the same time,  not every spatiotemporal pattern has something to do with the local bifurcations and connecting them is sometimes unclear or not likely as in the case of soliton-like waves reported in articles \cite{Ts03,{Ts04},{Ts04-1}}.  Nevertheless,  the instabilities of equilibria accompanied by the local bifurcations often are the first links in the chains of dynamical transitions leading to rather complex spatiotemporal patterns as it follows, e.g., from the observations reported in articles \cite{GMT,AGMTS,Chaplain,QiJiLu}.  The present article aims at studying the local stability issues for the counterparts of the equilibria in the systems driven by a signal produced externally. For example,  such a signal can be due to fluctuating the characteristics of environment such as temperature, salinity,  or the terrain relief.

Although considering the inhomogeneous PKS systems seems to be quite natural, mention-
ing them in the literature is much less often than mentioning the homogeneous ones. There are
several articles, e.g., \cite{Black1} or \cite{IssShn}  aimed at the topics like the global boundedness, extinction or coexistence but not at the issues we raise here. Also, we would like to mention article \cite{YrkCbbld} focused on the effect of the terrain relief on the excitation of waves in a spatially distributed living community. Although this article does not consider taxis, it employs homogenization, which we use too.

It turns out that the homogenized system also takes the form of the diffusion-advection system in which the advective flux consists of two different contributions,  both are emergent from the PKS cross-diffusion. The first one is the PKS cross-diffusion again, but this time it is driven by the averaged densities of the species, and the second one is the drift which is due to the external signal only. This drift is the main translator of the external signal effects. In particular, it is responsible for breaking the reflectional symmetry of the homogenized system  that   essentially influences the stability issues.

We examine the stability of the quasi-equilibria - that is, the simplest short-wavelength patterns fully imposed by the external signal. Comparing the results to the case of switching the signal off allows us to estimate the effect of it.  We give a detailed analysis  of the signals emitted as the travelling waves. The effect of such a wave on the stability of the quasi-equilibria turns out to be not single-valued but depending on the wave speed. Namely, there is an independent threshold value such that increasing the amplitude of the wave destabilizes the quasi-equilibria provided that the wave speed is above this value.  Otherwise,   the same action exerts the opposite effect. It is worth to note that the effect is exponential in the amplitude of the wave in both cases.

 The article consists of five sections supplemented with three appendices.  In section 1, we formulate the governing equations. In section 2, we address the short-wavelength external signals and describe the homogenized system.  In section 3, we introduce the equilibria and quasi- equilibria.   In section 4, we explore the stability and instabilities of the equilibria and quasi- equilibria. Section 5 contains the discussion on the obtained results. Appendices I, II and III contain the routine technical moments which regard the homogenization procedure, the stability analysis and asymptotics of some integral correspondingly.
\section{The governing equations}
\label{SecInrtTxs}
\setcounter{equation}{0}
\noindent
For a first attempt to exploring the effect of the external signals on the stability of quasi-equilibria,  we have picked out of the PKS family a system that is as simple as possible but still capable of forming the non-trivial spatiotemporal patterns due to the instabilities and local bifurcations of the homogeneous equilibria. Dimensionless form of this system reads as
\begin{eqnarray}
&& \pt_t\phi=q +\kappa^{-1}f-\nu \phi+\delta_u \pt^2_x\phi;
\label{SgnlEqPKS}\\
&& \pt_tp=\pt_x(\delta_p \pt_xp-\kappa p\phi_x);
\label{PrdEqPKS}\\
&& \pt_tq=q(1-q-p)+\delta_q \pt^2_x q.
\label{PryEqPKS}
\end{eqnarray}
Here $x,t$ stand for a spatial coordinate and time; $\partial_t$ and $\partial_x$ denote partial differentiation with respect to coordinates $t$ and $x$; $\delta_p,\delta_q,\delta_u$, $\kappa$,  $\nu$ denote the positive parameters.

Equations~(\ref{PryEqPKS}) and (\ref{PrdEqPKS}) describe   the balances of densities of two interacting species. At that, equation~(\ref{PrdEqPKS})   has got the PKS-term while equation~(\ref{PryEqPKS}) has not. In what follows, the species endowed (not endowed) with taxis stands as the predator  (prey).  We denote the predators' and prey' densities as $p$ and $q$, correspondingly. Note that the stimulus driving taxis is not the prey itself but another signal, intensity of which we have denoted  as $\phi$. Equation (\ref{SgnlEqPKS}) governs the production of the signal, which, in turn, gets an external contribution denoted as $f$.

It follows from equations (\ref{PryEqPKS}) and (\ref{PrdEqPKS}) that the reproduction and losses of the prey due to predation obey the logistic and  Lotka-Volterra laws correspondingly, and that the contribution from the reproduction and mortality of the predators is negligible.  The last assumption makes sense if  reproducing-dying the predators goes on much slower than the other processes considered.

Thus the prey-taxis modelled by system~(\ref{SgnlEqPKS})-(\ref{PryEqPKS})   is indirect in the sense that the signal is not the prey density but the intensity of some field emitted by prey.
Tello \& Wrozhek and also Li \& Tao have addressed such kind of taxis  in their recent articles \cite{TllWrzk} and \cite{LiTao} making focus upon the existence of the non-trivial steady states and  the global  boundedness of solutions. Tyutyunov et al. \cite{TtnZgr}, also recently,  have noticed the equivalence between the system~(\ref{SgnlEqPKS})-(\ref{PryEqPKS}) and  that introduced earlier in articles \cite{GMT,AGMTS} in an attempt of taking into  account the inertia of the species' transport. The latter system reads as
\begin{eqnarray}
&& \pt_tu=\pt_x(\kappa q+ f) -\nu u+\delta_u\pt^2_x u;
\label{TxEq}\\
&& \pt_tp=\pt_x(\delta_p \pt_xp-pu);
\label{PrdEq}\\
&& \pt_tq=q(1-q-p)+\delta_q \pt^2_x q.
\label{PryEq}
\end{eqnarray}
Bringing system~(\ref{SgnlEqPKS})-(\ref{PryEqPKS}) at the inertial form~(\ref{TxEq})-(\ref{PryEq}) employs a simple ansatz $u=\kappa\partial_x\phi$. It introduces new dependent variable denoted as $u$ which is nothing else than the velocity of  the predators advection. Equation~(\ref{TxEq}) governs this velocity  in response to the prey density and the external signal. Also, it takes into account the velocity diffusion and the resistance to the predators motion due to the environment. The coefficients of the diffusion and resistance are $\delta_u$ and $\nu$  correspondingly. The coefficient denoted as $\kappa$ stands as the measure of the prey-taxis intensity.

We pay attention to the formulated equivalence because the homogeneous version of the inertial system~(\ref{TxEq})-(\ref{PryEq}) (in which $f=0$) already had been  studied by Govorukhin et al. and Arditi et al. in above-cited articles, and they had reported the  transitions to the complex wave motions by the destabilization of the  homogeneous equilibria. We'll be considering only the inertial system~(\ref{TxEq})-(\ref{PryEq}) henceforth.
\section{Homogenization and drift}
\label{SecMgnDrft}
\setcounter{equation}{0}
\noindent
In what follows, let's consider fast variables $(\xi,\tau)$ as coordinates on 2-torus $\mathbb{T}^2$.
Define
\begin{equation}\label{DfnAvr}
    \langle g\rangle=\frac{1}{4\pi}\int\limits_{0}^{2\pi}\int\limits_{0}^{2\pi} g(x,t,\xi,\tau)\,d\xi d\tau.
\end{equation}
Let the external signal in  equation~(\ref{TxEq}) be a short wave, i.e.
\begin{equation}\label{ShrtWvSnl}
 f=f(x,t,\xi,\tau),\ \xi=\omega x,\ \tau=\omega t,\ \omega\gg 1
\end{equation}
and let the diffusion rates in equations~(\ref{TxEq}-\ref{PrdEq})  be of the same order as the  wave length, namely:
\begin{equation}\label{Dff2bSmll}
\delta_u=\nu_1\omega^{-1},\ \delta_p=\nu_2\omega^{-1},\ \nu_1=\const>0,\ \nu_2=\const>0.
\end{equation}
We state that under assumptions~(\ref{ShrtWvSnl}) and (\ref{Dff2bSmll}) the shortwave asymptotics of  system~(\ref{TxEq})-(\ref{PryEq})  takes the following form
\begin{eqnarray}
& q(x,t)=\bar{q}(x,t)+O({\omega}^{-1}),\ \omega\to+\infty;&
\label{q=}\\
&u(x,t)=\bar{u}(x,t)+\tilde{u}_0(x,t,\tau,\xi)+O({\omega}^{-1}),\  \omega\to+\infty;&
\label{u=}\\
& p(x,t)=\bar{p}(x,t)P(x,t,\tau,\xi)+O({\omega}^{-1}),
\ \omega\to+\infty; &
\label{p=}\\
&\partial_\tau \tilde{u}_0=\partial_\xi (f +\nu_1\partial_\xi \tilde{u}_0);\  \langle  \tilde{u}_0\rangle=0,&
\label{Tldu0Eq}\\
&\partial_\tau P=\partial_\xi(\nu_2\partial_{\xi}P -P(\bar{u}+\tilde{u}_0)),\ \langle  P\rangle=1;&
\label{PEq}\\
& \partial_t\bar{u}=\partial_x(\kappa \bar{q}+ \bar{f})-\nu \bar{u},\ \bar f=\langle f \rangle; &
\label{EqUSlw0}\\
 & \partial_t \bar{p}+\partial_x(\bar{p}(\bar{u}+\langle \tilde{u}_0P\rangle))=0;&
\label{EqPSlw0}\\
&\bar{q}_t=\bar{q}(1-\bar{p}-\bar{q})+\delta_q\partial^2_x\bar{q},&
\label{EqQSlw0}
\end{eqnarray}
where problems~(\ref{Tldu0Eq})-(\ref{PEq}) have to be solved on $\mathbb{T}^2$.  Deriving this approximation~(more or less routine) is  placed into Appendix I.

Specifying the external signal in the form of (\ref{ShrtWvSnl}) fully specifies equation~(\ref{Tldu0Eq}) which determines the shortwave velocity denoted as $\tilde{u}_0$,  which, in turn, enters the equation~(\ref{PEq})   as a coefficient. Then resolving the problem~(\ref{PEq}) and calculating  the mapping $\bar u\mapsto \langle \tilde{u}_0P\rangle$  puts equation~(\ref{EqPSlw0})  into a form involving only unknowns $\bar{p}$ and $\bar{u}$. Hence   equations~(\ref{EqUSlw0}), (\ref{EqPSlw0}) and (\ref{EqQSlw0}) form a closed system relative to unknowns $\bar{p},\bar{q}$, and $\bar{u}$ for every specific external signal.  This system is called \emph{ homogenized} henceforth.

Equation~(\ref{EqPSlw0}) which is responsible for the averaged transport of the predators  shows that their averaged velocity denoted as $\bar{u}$ is not the actual velocity of their advection -- that is, there is some drift, the velocity of which is equal to  $\langle \tilde{u}_0P\rangle$.  The drift velocity collects the ‘shortwave remembrances’ and passes them to the homogenized system.

 Let $f$ be given, we call the mapping $\mathcal{V}(f):\bar u\mapsto \langle \tilde{u}_0P\rangle$  \emph{the  drift operator}.  We'll be using the following notations and auxiliary notions. Let $\mathcal{G}:f\mapsto \tilde{u}_0$  be the transformation resulted from resolving  problem~(\ref{Tldu0Eq}). Further,  let $\sigma=\const\in\mathbb{R}$ and smooth function $\tilde{u}_0=\tilde{u}_0(x,t,\xi,\tau)$ be given, and, moreover, let $\langle\tilde{u}_0 \rangle=0$. We define the mapping
\begin{equation}\label{PMpng}
   \mathcal{P}(\tilde{u}_0):  \sigma\mapsto \langle\tilde{u}_0 P\rangle,
\end{equation}
where $P=P(x,t,\xi,\tau)$  is solution to  problem
\begin{equation}\label{SgmToP}
    \partial_\tau P=\partial_\xi(\nu_2\partial_{\xi}P -P(\sigma+\tilde{u}_0)),\quad \langle  P\rangle=1.
\end{equation}
Problem~(\ref{SgmToP}) has a unique solution for every $\sigma, \tilde{u}_0$ (by lemma proved in Appendix I). Once function $\tilde{u}_0$ is prescribed, mapping (\ref{PMpng}) parameterizes a path in  suitable space of functions in variables $(x,t)$. Finally, we arrive at the identity
\begin{equation}\label{Drft}
\left(\mathcal{V}(f)\bar{u}\right)(x,t)=\mathcal{P}(\mathcal{G}f)\bar{u}(x,t)
\end{equation}
In fact, the short-wavelength part  of the signal  determines the drift in full-- that is,
\begin{equation}\label{DrftAnd TldF}
\mathcal{G}f=\mathcal{G}\tilde{f},\ \ \mathcal{P}(\mathcal{G}f)=\mathcal{P}(\mathcal{G}\tilde{f}),\ \  \mathcal{V}(f)=\mathcal{V}(\tilde{f}),\ \ \tilde{f}=f-\langle f\rangle.
\end{equation}
Therefore, we write  the drift operator as $\mathcal{V}(\tilde{f})$, $\tilde{f}=f-\langle f\rangle$, instead of $\mathcal{V}({f})$ henceforth.
\begin{Exmpl}
\label{ExmplDrftTrWv}
Let us consider a short travelling wave $f(x,t,\eta)=\bar{f}(x,t)+A\tilde{f}(x,t,\eta)$, where $\eta=\xi-c\tau$, $A=\const\ge 0$, $c=\const\ge 0$, and function $\tilde{f}(x,t,\cdot)$ is $2\pi$-periodic and equal to zero on average for every $(x,t)$.\footnote{While considering the short-wavelength limit for these waves, we require $2\pi/c$-periodicity in $\tau$ instead of  $2\pi-$periodicity.   In accordance with this, we re-define the averaging, $\langle\cdot\rangle$. }
We'll be using the following notation. Let   $\partial_\eta^{-1}$ stand for the right inverse to   differentiation $\partial_\eta$  -- that is,
\begin{equation}\label{DefInvDxi}
    \partial_\eta \partial_\eta^{-1} w=w,\ \int\limits_0^{2\pi}\partial_\eta^{-1} w\,d\eta=0\quad \forall\ w:\ w(\cdot,\eta,\cdot)=w(\cdot,\eta+2\pi,\cdot),\ \int\limits_0^{2\pi}w\,d\eta=0.
\end{equation}.
Let  $g\ast h$ denote the common convolution of functions $g$ and $h$ on the real axis. Define
\begin{eqnarray}
 & \exp_\pm(\sigma)=\left\{\begin{array}{c}
                            \mathrm{e}^{\sigma},\ \pm \sigma>0, \\
                            0,\ \mp \sigma<0,
                          \end{array}\right.\label{DefExfPlsMns}\\
                          & s(x,t,\cdot)=-\pr^{-1}_\eta\exp_{+}^{-\frac{c}{\nu_1}}\ast \tilde{f}(x,t,\cdot);&
 \label{TrWvS}
\end{eqnarray}
where the superscript indicates raising  to a power, and the convolution acts in variable $\eta$. Then
\begin{equation}\label{TrWvVlct}
\tilde {u}_0=A\mathcal{G}\tilde{f}=\nu_1^{-1}A\pr_{\eta} s
\end{equation}
Further, we put
\begin{eqnarray}
   & E(x,t,\cdot)=\mathrm{e}^{as},\ a=(\nu_1\nu_2)^{-1} A, &
   \label{TrWvE}\\
   & R(x,t,\sigma)=\left\langle\frac{E(x,t,\cdot)}{E(x,t,\cdot-\sigma)}\right\rangle,&
   \label{TrWvR}\\
   & z=\frac{c-\bar{u}(x,t)}{\nu_2}.&
 \label{TrWvz}
\end{eqnarray}
Let $z\neq 0$. The periodic solution to equation~(\ref{PEq}) has the form
\begin{eqnarray}
&P(x,t,z,\cdot)=
 {\Gamma^{-1}_\pm}(x,t,z) {E(x,t,\cdot)\left(\exp_\pm^{-z}\ast E^{-1}(x,t,\cdot)\right)}
,\quad \pm z>0;&
\label{TrWvP}\\
 & {\Gamma_\pm}(x,t,z)=\int\limits_{\mathbb{R}}\exp_\pm^{-z}(\sigma) R(x,t,\sigma)\,d\sigma,\  \pm z>0.&
 \label{TrWvGm}
   \end{eqnarray}
Using the Fourier series matches  two expressions shown in~(\ref{TrWvP}) one to another one. Indeed,  let $\hat{E}_k(x,t)$ and $\check{E}_k(x,t)$ be the Fourier coefficients of functions $E(x,t,\cdot)$ and $E^{-1}(x,t,\cdot)$ correspondingly.  Then
\begin{eqnarray}
&P(x,t,z,\eta)= {\Gamma^{-1}}(x,t,z)
  {E(x,t,\eta)\left(\check{E}_0(x,t)+z\sum\limits_{k\in \mathbb{Z}\setminus\{0\}}  \frac{\check{E}_k(x,t)\mathrm{e}^{ik\eta}}{z+ik}\right)},&
\label{TrWvPFrSr}\\
 &\Gamma(x,t,z)={(\check{E}_0\hat{E}_0)(x,t)+z\sum\limits_{k\in \mathbb{Z}\setminus\{0\}}  \frac{(\hat{E}_k^*\check{E}_k)(x,t)}{z+ik}}.&
 \label{TrWvGmFrSr}
   \end{eqnarray}
Given the formula~(\ref{TrWvGmFrSr}), we express the drift velocity  as follows
\begin{equation}\label{TrWvDrftVlct}
    \mathcal{V}(A\tilde{f})\bar{u}=\langle\tilde{u}_0P\rangle =-\nu_2z\Gamma^{-1}(x,t,z)
  {\sum\limits_{k\in \mathbb{Z}\setminus\{0\}}  \frac{ik(\hat{E}_k^*\check{E}_k)}{z+ik}}=\nu_2z(1-\Gamma^{-1}(x,t,z)).
\end{equation}
(Deriving the second equality in this chain uses  identity
$\sum\limits_{k\in \mathbb{Z}}  \hat{E}_k^*\check{E}_k=\langle E E^{-1}\rangle=1.$) Further, let $v$ stand for the total advective velocity involved in equation~(\ref{EqPSlw0}), hence,
\begin{equation}\label{DefTtlAdvVlct}
    v\stackrel{\mathrm{def}}{=}\bar{u}+\mathcal{V}(A\tilde{f})\bar{u}.
\end{equation}
By equality (\ref{TrWvDrftVlct}),
\begin{eqnarray}
&v=v(x,t,z)=c-\frac{z\nu_2}{\Gamma\left(x,t,z\right)}&
\label{TrWvTtlTrnsprtVlct0}
   \end{eqnarray}
Since
   $$
   \Gamma(x,t,z)=\pm z\Gamma_\pm(x,t,z),\ \pm z>0,
   $$
we get additional representations for the total advective velocity, namely
\begin{eqnarray}
&v(x,t,z)=c\mp\frac{ \nu_2}{\Gamma_\pm \left(x,t,z\right)}\ \pm z>0.&
\label{TrWvTtlTrnsprtVlctPlsMns}
   \end{eqnarray}
Expressions~(\ref{TrWvGm}), (\ref{TrWvGmFrSr}), (\ref{TrWvTtlTrnsprtVlct0}), (\ref{TrWvTtlTrnsprtVlctPlsMns})  give the analytic continuation to the advective velocity to some strip parallel to the real axis in the complex plane of variable  $z$. This conclusion agrees with remark~\ref{RmOnPrctr} at the end of Appendix I.
\end{Exmpl}
\begin{Remark}
\label{RmOnRsdDrft}
Formula~(\ref{TrWvTtlTrnsprtVlctPlsMns}) (where one has to put $z=c/\nu_2>0$) shows that the total advective velocity is not equal to zero even if the averaged velocity denoted as  $\bar u$  vanishes. Indeed,
\begin{eqnarray}
&v(x,t,c/\nu_2)=\mathcal{V}(A\tilde{f})0=c-\frac{\nu_2}{\Gamma_+ (x,t,c/\nu_2)}\stackrel{\mathrm{def}}{=}v_e(x,t),\ c>0.&
\label{TrWvRsdlDrft}
   \end{eqnarray}
Thus, a short travelling wave emitted as the external signal induces a residual drift determined by equality (\ref{TrWvRsdlDrft}).  Inspecting the case of $c=0$ with the use of formula~(\ref{TrWvTtlTrnsprtVlct0}) shows that $\mathcal{V}(A\tilde{f})0=0$ -- that is,  the stationary waves  produce no residual drifts.
\end{Remark}
\begin{Remark}
\label{RmOncTo0}
Let $a>0$ and $c>0$ be the characteristic amplitude and phase velocity of some signal emitted as a travelling wave. Straightforward evaluating the expression~(\ref{TrWvRsdlDrft}) for the values of  $c$ tending to zero or infinity and for the fixed values of the other variables involved therein shows that
\begin{equation}\label{DrftVlctCTo0Infty}
    v_e\to 0 ,\ c\to+0,\quad v_e\to0,\ c\to\infty.
\end{equation}
Consider now expression~(\ref{TrWvRsdlDrft}) for the value of $a$ tending to zero or to infinity and for the fixed values of all other  quantities. In addition, assume that $\tilde{f}(x,t,\cdot)\not\equiv 0$.  Then
\begin{equation}\label{DrftVlctATo0,+infty}
    v_e\to 0,\ a\to+0, \quad v_e\to c,\ a\to\infty.
\end{equation}
The first limit written down in~(\ref{DrftVlctATo0,+infty}) is obvious while the second is a result of estimating the integral (\ref{TrWvGm}) with the use of the Laplace method. There are more details of this issue in Appendix III.
\end{Remark}

\section{Equilibria and quasi-equilibria}
\label{SecRltvEql}
\setcounter{equation}{0}
\noindent
Henceforth, we call as quasi-equilibrium~(equilibrium)  a stationary solution to homogenized system~(\ref{EqUSlw0}-\ref{EqQSlw0}) (to exact system~(\ref{TxEq}-\ref{PryEq})) such that  $\bar{u}=0$  ($u=0$).  Note that the definitions require  uniform distributions of the species neither for the  quasi-equilibria nor for the equilibria.

Given a quasi-equilibrium,  formulae~(\ref{q=})-(\ref{p=}) determine the leading approximation to a special solution to the exact system displaying the short-wavelength patterns of the predators' density and velocity.  At that, the latter vanishes on average for every value of the slow variables while the former does not do so, and distributing the averaged values of it is consistent with that specified by the quasi-equilibrium.

In the considerations below, the quasi-equilibria and the related short-wavelength regimes are not distinguishable in effect. Therefore we'll be using the same name for both.

\begin{figure}[h]
\centering
\includegraphics[scale=0.40]{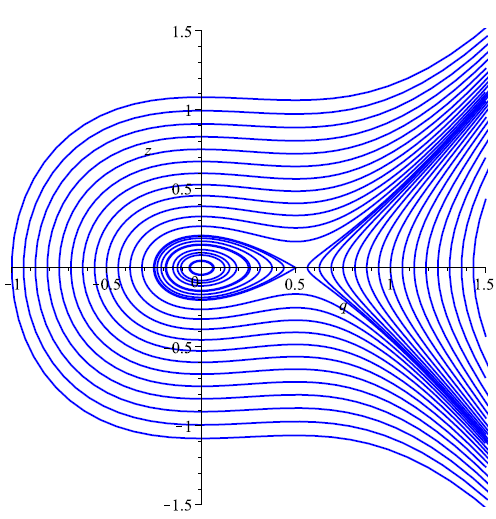}
\hfill\includegraphics[scale=0.40]{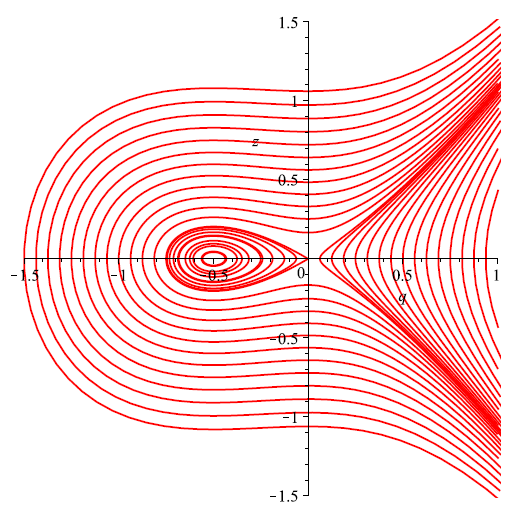}
\caption{\small Phase portraits of the first equation in system~(\ref{EqlbrEq}). The left frame displays $p=1/2$; the right frame displays $p=3/2$.}
\label{fig0}
\end{figure}

Tuning the external signal is necessary to set an equilibrium or quasi-equilibrium. Indeed, substituting the advective velocity, $u$, with zero in the exact system simplifies  it  as follows
\begin{equation}\label{EqlbrEq}
\delta_q\pr^2_{x}q+q(1-q-p)=0,\quad p=\const\ge 0,\quad \kappa q -\bar{f}=\const.
\end{equation}
Only the nonnegative solutions to problem (\ref{EqlbrEq}) make sense. However,  every such solution is unbounded  except for the constants. It is easy to see just looking at fig.~\ref{fig0}.   Hence $\bar{f}=\const$ too. Thus the translational invariance and homogeneous distributions of the species are necessary for bringing the exact system at equilibrium. In what follows, we neglect the trivial solution such that $q=0$ and arrive at homogeneous equilibria family
\begin{equation}\label{Eqlbr}
p\equiv p_e,\quad q\equiv q_e,\ u=0,\quad  p_e=\const>0,\quad  q_e=\const>0,\quad p_e+q_e=1.
\end{equation}
Now let's pass to the quasi-equilibria and consider them together with a bit more general motions of the homogenized system with zero mean velocity.  The equations governing such regimes are
\begin{eqnarray}
&\pr_t \bar{q}=\delta_q\pr^2_{x}\bar{q}+\bar{q}(1-\bar{q}-\bar{p}),\ \pr_t\bar{p}+\pr_x(v_e{\bar{p}})=0,\ v_e=\mathcal{V}(\tilde{f})0;\ \kappa \bar{q} -\bar{f}=\const,\ \bar f=\langle f\rangle&
\label{QEqlbrEq}
\end{eqnarray}
where $v_e$ is  the residual drift velocity. Conjecturing the existence of the suitable positive solutions to this problem links the smooth and short-wavelength parts of the signal one to another.  (These are $\bar{f}$ and $\tilde{f}=f-\langle f\rangle$). Exploring these interplays is far beyond the scope of this article, but we give a couple of examples.
\begin{Exmpl}
\label{ExmplInhmgEqlbr}
Let ${f}=\bar{f}+A\tilde{f}(x,t,\xi)$ where $A=\const\ge 0$, and let function $\tilde{f}(x,t,\cdot)$ be $2\pi$-periodic  and equal to zero on average for every $(x,t)$.  Then tending $z$ and $c$ to zero in formulae~(\ref{TrWvVlct}) and (\ref{TrWvPFrSr}) puts the quasi-equilibrium pattern into the following form
$$
\tilde{u}_0=\tilde{u}_e=-A(\nu_1\partial_\xi)^{-1}\tilde{f},\quad P=P_e={\langle\mathrm{e}^{as}\rangle}^{-1}\mathrm{e}^{as},\  a=(\nu_1\nu_2)^{-1}A,\ {s}=-\partial_\xi^{-2}\tilde{f}.
$$
Hence the signal does not produce the residual drift, i.e. $v_e=\mathcal{V}(\tilde{f})0=0$, as we have been noticing in remark~\ref{RmOnRsdDrft}. Hence  the equation for the averaged predators' transport reads as $\pr_t p=0$. Consequently, choosing a time-independent predators' density,  $p_e=p_e(x)$, determines the prey's density, $q=q_e(x)$, that, in turn, determines  signal's average, $\bar{f}=\bar{f}_e$. The positiveness restriction is easy to satisfy this time. In particular,  we arrive at the  homogeneous family (\ref{Eqlbr})  provided that  $\bar{f}_e=\const$.
\end{Exmpl}
\begin{Exmpl}
\label{ExmplHmgEqlbr}
Let $f=\tilde{f}(t,\xi,\tau)$ where $\langle\tilde{f}(t,\cdot)\rangle=0$ for every $t$. Then the spatial homogeneity of the residual drift velocity  is easy to see. Hence there exists the family of homogeneous quasi-equilibria~(\ref{Eqlbr}). Assuming that $f=A\tilde{f}(t,\eta)$, where $\eta=\xi-c\tau$, $A=\const\ge 0$, $c=\const\ge 0$, brings  us at the class of signals considered as example~\ref{ExmplDrftTrWv}, and we get explicit expressions for the quasi-equilibrium patterns and the residual drift velocity  by formulae~(\ref{TrWvVlct}), (\ref{TrWvP}), and  (\ref{TrWvRsdlDrft}) correspondingly by substituting  $z$ by $c/\nu_2$ in the two last ones. Note that the quasi-equilibrium patterns are the short travelling waves propagating at the same speed as those emitted externally.
\end{Exmpl}
\section{Stabilizing and destabilizing}
\label{SecStbltAndInst}
\setcounter{equation}{0}
\noindent
In what follows, we'll be investigating an effect of the short-wavelength signals on the stability of the quasi-equilibria by comparing them to the equilibria.  Such a comparison is most clear and natural in the case of homogeneous quasi-equilibria which we'll be restricted to henceforth. Note in passing, that only an external signal vanishing on average and producing a constant in space residual drift velocity denoted as $v_e$,  is capable of giving rise to a family of homogeneous quasi-equilibria~(\ref{Eqlbr}), and vice versa. For instance,  the  signal satisfying  both  conditions can take  the form  $f=A\tilde{f}(x,t,\xi)$  or $f=A\tilde{f}(t,\xi,\tau)$, where $\langle\tilde{f}\rangle=0$ for every value of the slow variable as  shown by examples~\ref{ExmplInhmgEqlbr} or \ref{ExmplHmgEqlbr}.

We start with revisiting the instability of the equilibria, which is known mainly due to the prior work by Govorukhin et al. and by Arditi et al., \cite{GMT},\cite{AGMTS}.
\subsection{Destabilizing the equilibria}
\label{SSecDstbEqlbr}
\noindent
We call as homogeneous the version of system~(\ref{TxEq})-(\ref{PryEq}) arising upon switching off  the external signal -- that is, for $f=0$. The homogenous system is invariant to the spatiotemporal translations. There exists the homogeneous family (\ref{Eqlbr}), as we have been mentioning in the section~\ref{SecRltvEql}.

Let's choose out of  family (\ref{Eqlbr}) an equilibrium with some specific densities $p_e>0$ and $q_e=1-p_e>0$.   The system governing  the evolution of a small perturbation of such an equilibrium  reads as
\begin{eqnarray}
&\pt_t {u}+\nu  {u}- {\kappa}\pt_{x} {q}=\delta_u\pt^2_{x}u,&
\label{TxEqLnr}\\
& \pt_t{p}+{p}_e \pt_{x} {u}=\delta_p \pt_{x}^2 {p};&
\label{PrdEqLnr}\\
& \pt_t {q}+q_e( {p}+ {q})=\delta_q \pt_{x}^2 {q};&
\label{PryEqLnr}\\
&  {p}_e+ {q}_e=1. &
\nonumber
\end{eqnarray}
Given the invariance of system~(\ref{TxEqLnr})-(\ref{PryEqLnr}) to the spatiotemporal translations,   we'll be considering only  the eigenmodes  having  the following form
\begin{equation}\label{EgnMds}
       (\hat{u},\hat{p},\hat{q})\exp(i\alpha x+\lambda t),\ \lambda=\lambda(\alpha)  \in \mathbb{C},\ \alpha\in \mathbb{R}.
\end{equation}
Here $\lambda$ is the  eigenvalue of the  spectral problem arising from  the substituting of eigenmodes~(\ref{EgnMds}) into  system~(\ref{TxEqLnr})-(\ref{PryEqLnr}).  We say that eigenmode~(\ref{EgnMds}) is stable~(unstable, neutral) if the real part of  the corresponding eigenvalue is negative (positive, equal to zero). We'll be looking for the occurrences of instability, that is,  transversal intersecting the imaginary axis by a smooth branch of eigenvalues upon changing  the other parameters of the spectral problem along a smooth path. If such a branch crosses the imaginary axis at a non-zero point, the instability is named as oscillatory otherwise as monotone.

It is well-known that an occurrence of instability in the family of equilibria of a smooth family of vector fields indicates the local bifurcations. If there are no additional degenerations, branching the equilibria family accompanies the monotone instability, and branching the limit cycle off the family accompanies the oscillatory instability  (Poincare-Andronov-Hopf bifurcation), and more complex bifurcations happen in the case of additional degeneracy, e.g. when the neutral spectrum is multiple. For more information on this subject, a reader could refer to monographs \cite{IssJsph,ArnAfIlSh,HrgsIss}.

\begin{figure}[h]
\centering
\includegraphics [scale=0.50]{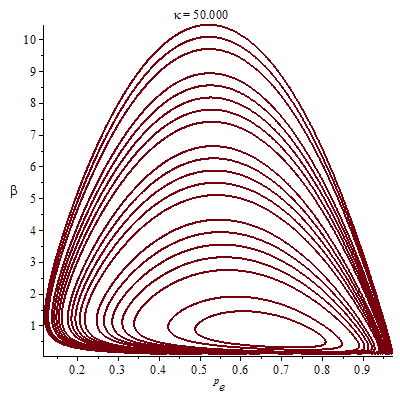}\hfill
\includegraphics [scale=0.50]{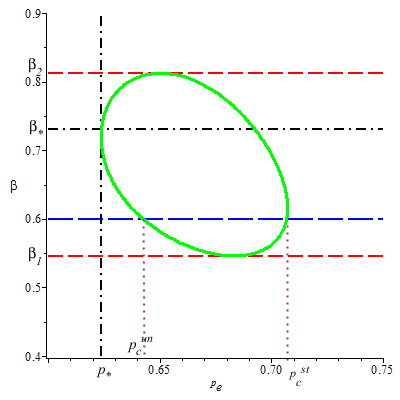}
\caption{\small  The {\sl left frame} shows typical curves determined by equation  $\kappa_c(p_e,\beta,{\delta})=\kappa$ in the semi-strip $\{0<p_e<1,\beta>0\}$ for the values of $\kappa$  in the range from $15.12$ to $50.0$  and ${\delta}=(1,1,0,0)$. The area bounded by such a curve increases with $\kappa$ and tend to cover all the semi-strip when $\kappa\to+\infty$. At that, $
\left.\kappa_*({\delta})\right|_{{\delta}=(1,1,0,0)}=27/2,
$
and $\kappa_c$ takes this value at $p_e=2/3$, $\beta=2/3$. The {\sl right frame} corresponds to ${\delta}=(1,1,0,0)$ and
$\kappa=13.56$. The other numerical values are: $p_*\approx 0.624$; $p_c^{un}\approx0.642$, $p^{st}_c\approx0.707$, $\beta_1\approx0.546$,  $\beta_2\approx0.812$. }
\label{fig1}
\end{figure}

It is convenient to introduce the following notation
$$
\beta=\alpha^2,\ \delta=(\nu,\delta_q,\delta_p,\delta_u).
$$
Note that each equilibrium~(\ref{Eqlbr}) has a neutral homogeneous mode (that corresponds to $\lambda=\alpha=0$), but this does not lead to any long-wave instabilities. To get rid of  this and  other unwanted degenerations, we assume the following
\begin{equation}\label{Restr}
\beta>0,\ 0<p_e<1,\    \nu(\delta_p+\delta_u+\delta_q)>0.
\end{equation}
Let $\Pi$ be a domain cut out by  inequalities~(\ref{Restr}) in the space of parameters $p_e,\beta,\delta$.  Let us consider  an equilibrium of family (\ref{Eqlbr}) with specific density $p_e$ and its  eigenmodes~(\ref{EgnMds}) with specific wavelength $\alpha=\pm\sqrt{\beta}$.   There exists function
$$
\kappa_c = \kappa_c (p_e, \beta,\delta)
$$
analytic in $\Pi$ and such that (i) each of those  eigenmodes  is stable provided that $\kappa<\kappa_c (p_e, \beta,\delta)$; (ii) there is an unstable mode  provided that $\kappa>\kappa_c (p_e, \beta,\delta)$; (iii) there exists two conjugated neutral modes with $\lambda\neq 0$ provided that $\kappa=\kappa_c(p_e, \beta,\delta)$.  The oscillatory instability  occurs  each time  a path in  $\Pi\times(0,\infty)$ (where $(0,\infty)\ni\kappa$) intersects  graph $\{(p_e,\beta,\delta),\kappa_c(p_e,\beta,\delta)\}$  transversally, perhaps, with except for some cases of degeneracy (Fig.~\ref{fig1}).  Note that
\begin{equation}\label{DfTrhKpp} \min\limits_{0<p_e<1,\,\beta>0}\kappa_c(p_e,\beta,\delta)=\kappa_*(\delta)>0
\end{equation}
where the strict positiveness  takes place for every $\delta$ obeying (\ref{Restr}). For every $\kappa>\kappa_*(\delta)$, equation $\kappa=\kappa_c(p_e,\beta,\delta)$ determines a closed curve inside semi-strip $\{0<p_e<1,\,\beta>0\}$. This curve widens itself and  tends to the boundary of the  semi-strip as $\kappa\to+\infty$.
Fig.~\ref{fig1} shows that the oscillatory instability of homogeneous equilibria occurs in response to increasing the specific predators' density denoted as  $p_e$ provided that the predators' motility is  above the threshold -- that is, provided that the inequality $\kappa>\kappa_*(\delta)$ holds. For more details on the stability analysis, a reader can refer to the first subsection of Appendix II.

By direct numerical simulations, Govorukhin et al. and Arditi et al. demonstrated that this instability leads to exciting the waves which are time-periodical for the weakly supercritical values of $p_e$, and which tend to move chaotically for the greater values of $p_e$. Also, it turned out that the wavy motions anyway allow the predators to consume more while keeping a greater stock of prey, and, in this sense, the waves are always more advantageous than the equilibria.


At this point, it is worth to stress the role of quantity $\kappa_* (\delta)$ defined in~(\ref{DfTrhKpp}). If $\kappa<\kappa_* (\delta)$ then neither the oscillatory instability of the homogeneous equilibrium nor the accompanying bifurcation is possible, whatever values of the predators' density and the disturbances wavelengths are specified.  In this sense, the value of $\kappa_* (\delta)$ is the threshold of the absolute stabilization of the homogeneous equilibria.  If the predators' motility is below this threshold for some predator-prey community, then such a community fails to adapt itself to the resource deficiency. It is important to note that $\kappa_* (\delta)>0$ for every $\delta$ obeying restriction~(\ref{Restr}).


The neutral spectrum we face upon the considered instability is always multiple.  Namely, the pure imaginary pair of eigenvalues is double, and there is a simple null eigenvalue.  For such a degeneration,  there are two reasons. The first is the reflectional symmetry. The second is the conservation law for the predators' density,  because of which the homogeneous equilibria are not isolated but form the continuous 1-parametric family.

In the mentioned studies of the homogeneous system, Govorukhin et al. and Arditi et al.  formulated the initial-boundary value problem for a bounded spatial domain with the Neumann's boundary conditions  (also known as the no-flux boundary conditions).   Such setting removes the reflectional symmetry together with the multiplicity of the pure imaginary pair, but the conservation law persists. The equilibrium family and the null eigenvalue persist too.  One can get rid of this residual degeneration by restricting the system on the level sets of the conserved quantity.  As a result,   the common theory of Poincare-Andronov-Hopf bifurcation becomes applicable to the restriction.  Alternatively, one can apply the general results on  the bifurcation accompanying the oscillatory instability for the vector fields, which possess the so-called cosymmetry \cite{YdCsmOscInst}\footnote{Given the cosymmetry,  the limit cycle generically does not branch off from the critical equilibrium except for integrable cosymmetry that is equivalent to a conservation law.    In case of such an exception,  the cycle branches off  `as usual' provided that there is no additional degeneration.}.

Recently, several authors \cite{Chaplain,TtnZgr,QiJiLu} explored the oscillatory instability and the spatiotemporal patterns it creates in more general but still homogeneous PKS systems. They considered the finite domains and employed the Neumann's boundary conditions.  Given such boundary conditions and the kinetic of species, the mentioned degenerations disappear, and the oscillatory instability follows the generic Poincare-Andronov-Hopf scenario.



\subsection{Destabilizing the  quasi-equilibria}
\label{SSecDstbQsEqlbr}
\noindent
Let the homogeneous quasi-equilibria  form family (\ref{Eqlbr}) for some specified signal.   Let's choose out of this family a quasi-equilibrium with some specific densities $p_e>0$ and $q_e=1-p_e>0$.   The system governing  the small perturbations of such a quasi-equilibrium has the form
\begin{eqnarray}
& \pt_t\bar{u}+\nu \bar{u}-\kappa  \pt_{x}\bar{q}=0;&
\label{TxEqLnrSlw0}\\
& \pt_t\bar{p}+\pt_{x}(v_e\bar p+ {p}_e(1+\mathcal{V}^\prime(\tilde{f}))\bar{u})=0;&
\label{PrdEqLnrSlw0}\\
& \pt_t\bar{q}+{q}_e(\bar{p}+\bar{q})-\delta_q \pt^2_{x}\bar{q}=0,&
\label{PryEqLnrSlw0}
\end{eqnarray}
where  $v_e=\mathcal{V}(\tilde{f})0$, and $\mathcal{V}^\prime(\tilde{f})$ stands for  the differential of the drift operator evaluated at $\bar{u}=0$.   By definitions~(\ref{Drft}) and (\ref{PMpng}), $\left(\mathcal{V}^\prime(\tilde{f})w\right)(x,t)=\left(\mathcal{P}^\prime(\mathcal{G}\tilde{f})\right)w(x,t)$, where  $\mathcal{P}^\prime(\mathcal{G}\tilde{f})$ denotes the differential of the mapping $\mathcal{P}(\mathcal{G}\tilde{f}):\mathbb{R}\to \mathbb{R}$  evaluated at the origin. As long as there are no additional  assumptions regarding the signal, the mapping $\mathcal{P}(\mathcal{G}\tilde{f})$ itself depends on $(x,t)$. Hence we identify the action of $\mathcal{V}^\prime(\tilde{f})$ with multiplying by certain real-valued function in variables $(x,t)$. Namely,
\begin{equation}\label{DffV}
    \mathcal{V}^\prime(\tilde{f})=
    \left.\frac{d}{d\sigma}\right|_{\sigma=0}\mathcal{P}^\prime(\mathcal{G}\tilde{f})(\sigma)
    =\left.\frac{d}{d\sigma}\right|_{\sigma=0}\langle\tilde{u}_0P\rangle=\langle  \tilde{u}_0P_1\rangle
\end{equation}
 where  $P_1$ is determined by equations
\begin{eqnarray}
   &\pr_\tau P_1= \partial_\xi(\nu_2\partial_{\xi}P_1 -P_e-\tilde{u}_0 P_1),\ \langle  P_1\rangle=0,\quad P_e=\mathcal{P}(\mathcal{G}\tilde{f})0&
 \label{P1onXi}  
\end{eqnarray}
The signals free of slow modulating, i.e. determined by functions $f=\tilde{f}(\xi,\tau)$, produce the drift operators and the homogenized systems, which are invariant to spatiotemporal translations.  Given such an invariance, we are capable of comparing the stability of the equilibria to the quasi-equilibria in the most direct manner.   In particular, $v_e=\mathcal{V}(\tilde{f})0=\const$, and  the action of $\mathcal{V}^\prime(\tilde{f})$ is nothing else than multiplying by a real constant. Hence the coefficients of the system~(\ref{TxEqLnrSlw0})-(\ref{PryEqLnrSlw0})  become constant as well as in the case of the linearized homogeneous system, and  we can again examine the stability of quasi-equilibria using only eigenmodes~(\ref{EgnMds}).
\begin{Remark}
\label{RmOnCnstDrftVlc}The residual drift makes sense despite the constant velocity since there is a distinguished coordinate system, relative to which the external signal is specified. For instance, the signal emitted as a travelling wave distinguishes the coordinate system relative to which the speed it propagates at  takes the prescribed value.
\end{Remark}


Let's narrow the class of signals to short unmodulated travelling waves defined by functions
\begin{equation}\label{TrWvUnMdltd}
    f=A\tilde{f}(\eta),\ \eta=\xi-c\tau,\ A=\const\ge 0,\ c=\const\ge 0.
\end{equation}
Thus, the coefficients of system~(\ref{TxEqLnrSlw0})-(\ref{PryEqLnrSlw0}) are constant.   Re-scaling the unknown denoted as  $\bar u$ puts the system into the form
\begin{eqnarray}
& \pt_t\bar{u}+\nu \bar{u}+\bar{\kappa}  \pt_{x}\bar{q}=0;\bar\kappa=(1+\mathcal{V}^\prime(\tilde{f}))\kappa&
\label{TxEqLnrSlw}\\
& \pt_t\bar{p}+\pt_{x}(v_e\bar p+ {p}_e\bar{u})=0;&
\label{PrdEqLnrSlw}\\
& \pt_t\bar{q}+{q}_e(\bar{p}+\bar{q})-\delta_q \pt^2_{x}\bar{q}=0.&
\label{PryEqLnrSlw}
\end{eqnarray}
In what follows, the coefficient denoted as $\bar\kappa$ in equation (\ref{TxEqLnrSlw}) is called the effective motility .

System~(\ref{TxEqLnrSlw})-(\ref{PryEqLnrSlw}) is   similar but not identical to the linearization of the exact system given by equations~(\ref{TxEqLnr})-(\ref{PryEqLnr})   where $\delta_p=\delta_u=0$, and $\kappa$  changed by $\bar\kappa=(1+\mathcal{V}^\prime(\tilde{f}))\kappa$. The essential difference is due to  the residual drift velocity, $v_e$. Therefore, the effect of the short-wavelength external signal on the linear stability of quasi-equilibria consists in producing the residual drift and altering the prey-taxis intensity or, equivalently, the predators' motility by transformation $\kappa\mapsto(1 +\mathcal{V}^\prime(\tilde{f}))\kappa$.

If the residual drift vanishes  then equations~(\ref{TxEqLnrSlw})-(\ref{PryEqLnrSlw}) becomes   identical to those listed in~(\ref{TxEqLnr})-(\ref{PryEqLnr}) (where $\delta_p=\delta_u=0$, and  $\kappa$  changed by $\bar\kappa=(1+\mathcal{V}^\prime({f}))\kappa$). Let's assume for a while that the signal takes the form
\begin{equation}\label{FonXi}
{f}=A\tilde{f}(\xi),\  \langle{\tilde{f}}\rangle=0.
\end{equation}
As we have been mentioning while discussing example~\ref{ExmplInhmgEqlbr},  such a signal  does not produce a residual drift -- that is, $v_e=0$.  Since restrictions~(\ref{Restr}) imposed in Sec.~\ref{SSecDstbEqlbr} on the problem parameters allow us to null  the diffusion rates $\delta_p$ and $\delta_u$ simultaneously,  all we have been saying   in Sec.~\ref{SSecDstbEqlbr} about the   stability of homogeneous equilibria  is  also true regarding  the quasi-equilibria  upon replacing the predators motility, $\kappa$,   by its effective counterpart $\bar{\kappa}=(1+\mathcal{V}^\prime({f}))\kappa$. In particular,  every  eigenmode  of  every quasi-equilibria is stable  provided that the following inequality  holds
\begin{equation}\label{NewAbsTrhld0}
1+\mathcal{V}^\prime({f})<{\kappa}^{-1}{\kappa_*|_{\delta_p=\delta_u=0}},
\end{equation}
where $\kappa_*$ is exactly the threshold motility of the predators defined by equality (\ref{DfTrhKpp}) in Sec.~\ref{SSecDstbEqlbr}.  The right-hand side in inequality (\ref{NewAbsTrhld0}) does not depend on the external signal while the left-hand side does depend.  Hence it makes sense to ask whether manipulating the external signal can switch the values of inequality (\ref{NewAbsTrhld0}) from false to true. Such switching eliminates every instability no matter which quasi-equilibria and which perturbation's wavenumber we consider, and in this sense the achieved stabilizing is absolute.

Let's show that the answer to the raised question is affirmative.   Given the definition by equality (\ref{DefTtlAdvVlct}), the total advective velocity denoted as $v$ represents the image of the mapping $\bar{u}\mapsto \bar{u}+\mathcal{V}(A\tilde{f})\bar{u}$. Hence the action of the differential of this mapping evaluated at  $\bar{u}=0$  is nothing else than multiplying by the effective motility factor equal to
$
 1+\mathcal{V}^\prime({f}).
$
At the same time, equality (\ref{TrWvTtlTrnsprtVlct0}) gives explicit expression to the total advective velocity. Differentiating it in variable $z$ at $z=0$ while keeping in mind assumption (\ref{FonXi})leads to equality
\begin{equation}\label{SttnrWvMtltFctr}
    1+\mathcal{V}^\prime({f})=
 \frac{1}{\langle\mathrm{e}^{-as}{\rangle}\langle\mathrm{e}^{as}{\rangle}},\ \ a=\frac{A}{\nu_1\nu_2},\ s={-\partial_\xi^{-2}{\tilde{f}}}.
\end{equation}
 Replacing the left-hand side in inequality (\ref{NewAbsTrhld0}) by the right-hand side of the last  equality brings us at an explicit formulation of the criterion for the absolute stabilizing that  reads as
\begin{equation}\label{NewAbsTrhld}
\frac{1}{\langle\mathrm{e}^{-as}{\rangle}\langle\mathrm{e}^{as}{\rangle}}
<\kappa^{-1}\kappa_*|_{\delta_p=\delta_u=0}.
\end{equation}
\begin{Exmpl}
\label{Exmplsn}
Let $f=A\sin\xi$, $A=\const>0$. Then $s=\sin\xi$, and
$$
\langle\mathrm{e}^{-as}\rangle\langle\mathrm{e}^{as}{\rangle}=I^2_0(a)
$$
where $I_0$ is the modified Bessel function of first kind. Consequently,  the criterion~(\ref{NewAbsTrhld}) for the absolute  stabilization takes the following form
\begin{equation}\label{NewAbsTrhldSn}
    I^{-2}_0(a)<{\kappa}^{-1}\kappa_*|_{\delta_p=\delta_u=0}.
\end{equation}
The quantity denoted as $a$ represents a characteristic amplitude of the external signal. When this amplitude grows up,  the left-hand side in inequality (\ref{NewAbsTrhldSn}) goes down exponentially as well as the effective motility, therefore. Hence the increase in the level of the external signal yields almost immediately the absolute stabilization of the quasi-equilibria due to the exponential losses  in the predators' motility.
\end{Exmpl}
\begin{Remark}
\label{Rmrk2} The exponential loss of motility discovered in example~\ref{Exmplsn} is almost independent on the specific form of the signal provided that it is stationary in the sense of the restriction~(\ref{FonXi}).   At such a conclusion, we arrive by estimating the averaged values involved in expression~(\ref{NewAbsTrhld}) for $a\to\infty$ by the Laplace's method (see Appendix III for more details). Note in passing, that the effect takes place irrespective of whether the signal is an attractant or repellent for the predators.
\end{Remark}
Now we get rid of restriction~(\ref{FonXi}) and proceed with the short travelling waves introduced by equality  (\ref{TrWvUnMdltd}). Formula~(\ref{TrWvRsdlDrft}) shows that the residual drift velocity takes a non zero value for every $c\neq 0$. Putting $c=0$ brings us back to the case of stationary signals.

\begin{figure}[h]
\centering
\includegraphics
[scale=0.50]{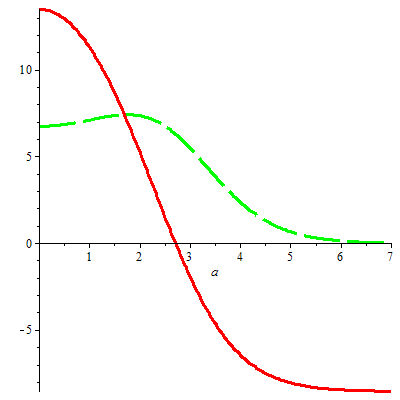}\hfill\includegraphics [scale=0.50]{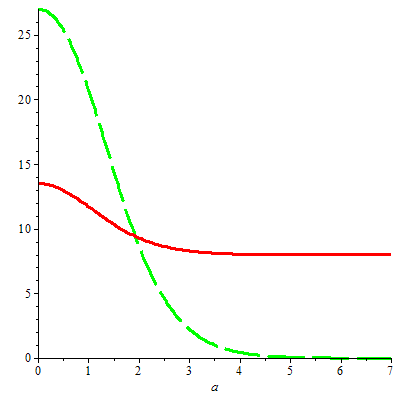}
\caption{\small The  figure displays the plots $\kappa_c^{-}$ vs $a$   and $\bar\kappa$ vs $a$ drawn by the solid and long-dashed lines correspondingly. For computing them, we have been applying the explicit expressions derived in example~\ref{ExmplDrftTrWv} to the specific signal profile $\tilde{f}(\eta)=\sin\eta$. The right (left) frame displays the plots drawn for $c=1/2$ ($c=3/2$). The case of homogeneous system corresponds to $a=0$. The values of the other parameters are as follows: $\delta=\nu=1$, $p_e=\beta=2/3$. For such parameters, $c_*=\sqrt{3/2}$ and $\kappa_c=\kappa_*=27/2$. Thus, $c>c_*$ ($c<c_*$) for the plots shown in the left (right) frame. For $a\to+0$, the limit value of $\bar\kappa$ is equal to $\kappa=2\kappa_*=27$ ($\kappa=\kappa_*/2=27/4$) for the right frame (left) frame.  From what we have been saying above regarding stability and instability of the quasi-equilibria, it follows that the eigenmode with squared wavenumber $\beta=2/3$  associated to the quasi-equilibrium with the value of predator's density equal to $p_e=2/3$   is unstable (stable) for every value of  $a$ for which long-dashed line is above (below) the solid one. Hence  we see in the left frame that the growth of $a$ destabilizes the eigenmode, which is stable for the values of $a$ close to zero.  Further increase in $a$ entails the occurrence of the negative values of $\kappa_c^{-}$ (that causes the total destabilizing). In the right frame, we see that the growth of $a$ stabilizes the eigenmode, which is unstable for the values of $a$ close to zero.
}
\label{figDstb-Stb}
\end{figure}

It follows from the stability analysis that the residual drift substantially rearranges the critical submanifolds in the space of the problem parameters in comparison to those reported for the case of a stationary signal. The main difference is that the oscillatory instability generically occurs twice upon crossing the    graphs of functions $\bar{\kappa}=\kappa_c^\pm(v_e,p_e,\beta,\delta)$ defined on  the intersection of hyperplane $\{\delta=(\nu,\delta_q,0,0)\}$ with the domain introduced and denoted as $\Pi$ in subsection~\ref{SSecDstbEqlbr}. (Imposing the additional restriction on $\delta$ is due to the zero values of  $\delta_u$ and $\delta_p$.)    Specific eigenmode~(\ref{EgnMds}) having wavenumber $\alpha=\pm\sqrt{\beta}$ is stable provided that the following inequality holds
\begin{equation}\label{TrWvStblArea}
    (1+\mathcal{V}^\prime(f))\kappa<\kappa_c^-(v_e,p_e,\beta,\delta).
\end{equation}
 A reader can find more details in the second subsection of Appendix II. It follows from the expressions  for the values of $\kappa_c^\pm$ given by
equality (\ref{KppaCrPM}) of  Appendix II that
\begin{equation}\label{Kppc-<Kppc}
     \kappa_c^-(v_e,p_e,\beta,\delta)<\kappa_c^-(0,p_e,\beta,\delta)=\kappa_c(p_e,\beta,\delta)\ \forall v_e\neq 0.
\end{equation}
Moreover,  increasing the residual drift velocity decreases the value of $\kappa_c^-$ and destabilizes, therefore.  Interestingly, the values of $\kappa_c^-$ become negative for
\begin{equation}\label{DefC*}
  v_e>  c_*=\sqrt{\frac{\nu(q_e+\delta_q\beta)}{\beta}}
\end{equation}
In such an occasion, inequality  (\ref{TrWvStblArea}) necessarily fails to hold, and therefore, the considered eigenmode is unstable for every admissible set of the problem parameters.  The answer to the question whether varying the amplitude of the travelling wave that determines the signal can give rise to this destabilizing is not single-valued but depending on the relation of the wave speed to the independent threshold value determined in (\ref{DefC*}). The left and right frames in  Fig.~\ref{figDstb-Stb} illustrate the possibilities arising for $c>c_*$ and $c<c_*$  correspondingly.

Let $c>c_*$. Then it follows from the limit identities~(\ref{DrftVlctATo0,+infty}) that $v_e>c_*$ and $\kappa_c^-<0$ provided that the value of $a$ is sufficiently high, and the opposite inequalities hold provided that the value of $a$ is sufficiently small.  Thus, for $c>c_*$,  increasing the value of $a$  leads to the total destabilization while decreasing it leads to certain stabilization.  This conclusion sharply contrasts to that reported for the case of the stationary signals.

Let now $c<c_*$. Then the right-hand side of inequality (\ref{TrWvStblArea}) is positive for sufficiently high values of $a$ while the effective motility factor on the left-hand side is equal to the first derivative of expression~(\ref{TrWvTtlTrnsprtVlctPlsMns}) in variable $z$ at $z=c/\nu_2$ -- that is,
\begin{equation}\label{TrWvMtltFctr}
 1+\mathcal{V}^\prime(f)=\frac{1}{\Gamma_+^2(\frac{c}{\nu_2})}\int\limits_0^\infty R(\sigma){\mathrm{e}^{-\frac{c\sigma}{\nu_2}}\sigma}\,d\sigma =\left.\frac{d\Gamma_+^{-1}(z)}{dz}\right|_{z=\frac{c}{\nu_2}}.
\end{equation}
 The right-hand side in equality (\ref{TrWvMtltFctr}) again decays exponentially when $a\to\infty$, see Appendix III for details. Hence increasing  the  value of $a$ stabilizes the quasi-equilibria in the same way as in the case of the  stationary signal.
\begin{Remark}
\label{RmOnShrtWvDstbl}
Upon increasing the drift velocity,  destabilizing the eigenmodes begins from the `shortwave end of the spectrum'. Indeed, inequality (\ref{DefC*})  holds for every eigenmode the wavenumber of which satisfies the inequality.
\begin{equation}\label{DefBta*}
   \beta>\beta_*=\frac{q_e\nu}{v_e^2-\nu\delta_q}.
\end{equation}
\end{Remark}
\section{Concluding remarks}
\label{SecCnclsn}
\setcounter{equation}{0}
\noindent
We have been seeing exponential suppressing the predators' motility in response to the growth of the signal amplitude. The natural treatment of this observation is that perceiving the intensive external signals distracts the predators from pursuing the prey. Given such an explanation, the effect might seem rather predictable, yet the exponential acuity of it remains surprising.

Intuitively, the loss of the predators' motility could be enhancing the stability of the homogeneous quasi-equilibria and suppressing the excitation of waves but, generally speaking, this is not the case, see fig.~\ref{figDstb-Stb}. Indeed,     increasing the amplitude of the external signal represented by a travelling wave gives rise to the total instability provided that the external wave speed is above the independent threshold defined by equality (\ref{DefC*}).  Otherwise,   the same action exerts the stabilizing effect, that becomes most pronounced in the limit case of the stationary waves.

In the end, increasing the amplitude of the external wave brings the system at almost degenerated form due to losing the motility. At that, the residual drift velocity becomes almost equal to that of the external wave. (Using the word `almost' here indicates  allowing the exponentially small deviations.)  In the linear approximation,  the degeneration reveals itself by the almost neutral waves propagating for every wavenumber at the phase velocities almost equal to the external wave speed.     The instability relative to such waves seems to be very weak. However,  they can interact non-linearly with the developed regimes which had been created by the local bifurcation at the smaller amplitude. (Imagine the threshold value of the motility decreasing and passing through  the  current value of the motility upon increasing the amplitude.)

The short-wavelength stabilization or destabilization of the quasi-equilibria resemble the effects of high-frequency vibration widely known in classical mechanics or the continuous media mechanics.   The celebrated examples are the upside-down pendulum or the counterparts of it emergent from the dynamics of the stratified fluid. The vibrations translate their effect on such kind of systems via so-called effective potential energy, that arises upon averaging, see, e.g., articles by \cite{Yudovich,Vldmrv}.  Our examples show that the short-wavelength fluctuations of the environment exert their effects on the PKS systems quite differently, namely, by adding the drift.   Studying the drift arising upon averaging the advection of some density by an oscillating velocity goes back to Stokes. Now this area is the subject of continuing  researches. A reader can find more details and references, e.g., in the article by \cite{Vldmrv1}.  Usually,  Stokes' drift turns out to be relatively weak, and the effect of it reveals itself only regarding the long-term transport or mixing. Nevertheless, we have been observing stabilizing or destabilizing the quasi-equilibria only due to  Stokes's drift right in the leading approximation.

In our considerations, the unstable modes occur upon crossing the graphs of two critical values of motility denoted as $\kappa_c^\pm$. Right on each graph, there exists a neutral wave. The neutral wave occurring at the lower critical value equal to $\kappa_c^-$ propagates upstream, i.e. faster than the residual drift, and the other one propagates downstream, i.e. slower than the residual drift. Since the value of $\kappa_c^-$ is the stability threshold, the wave propagating upstream destabilizes the quasi-equilibria.  When the residual drift vanishes,  the upper and lower critical motilities collide and merge with that reported in the case of equilibria. These features are due to breaking the reflectional symmetry by the drift. In particular,   the double neutral mode existing due to the reflectional symmetry splits itself into two simple ones. Thus we arrive at a two-parametric phenomenon that deserves detailed nonlinear analysis.

Considering the PKS systems in the bounded spatial domains allows us to avoid some difficulties emergent from the continuous spectra, but brings us at the problem of choosing the boundary conditions.  For instance, imposing the Neumann conditions as done in the cited articles leads to some technical issues upon homogenizing the problem and even upon doing the linear analysis. Addressing these technicalities does not seem to be essential since it is not clear whether the Neumann conditions are less artificial than the others. Besides, reducing the symmetry by the boundary conditions can cut out some solutions.  In such circumstances, the softer conditions of spatial periodicity seem to be the best. Under such conditions, the homogeneous system~(\ref{TxEq})-(\ref{PryEq}) and more general PKS systems possess   the translational and mirror symmetries. Although it doubles the neutral modes, the corresponding bifurcations can be addressed using the general theory given in article \cite{YudMrsh}.

Two more apparent topics for continuing the present research are the short standing waves and the slowly modulated travelling waves.  For instance, the functions $f=A\sin\tau\sin\xi$ and $f=A\sin t\sin(\xi-c\tau)$  stand as the representatives of the former and the latter kinds correspondingly. Both classes of signals produce the homogeneous quasi-equilibria but the coefficients of the homogenized system become variable. In the case of time-periodic slow modulation, one can address the corresponding spectral stability problem using Floquet's theory.

Finally, it is of interest to what extent the shape of the external signal influences its effect. In particular, there arises an interesting   optimization problem for the effective motility defined in~(\ref{SttnrWvMtltFctr}) or in~(\ref{TrWvMtltFctr}) subject to restriction
$
\langle f^2\rangle=1.
$



\subsubsection*{Acknowledgments} Andrey Morgulis acknowledges the support from Southern Federal University (SFedU)

\section{Appendix I. Derivation of the asymptotics }
\setcounter{equation}{0}
\noindent
In this appendix, we derive formally the asymptotic approximation described in Sec.~\ref{SecMgnDrft}.

Introducing the fast variables $\xi=\omega x$, $\tau=\omega t$ into the governing equations~(\ref{TxEq})-(\ref{PryEq}) puts them into the following form
\begin{eqnarray}
  &\omega((\partial_t+\omega\partial_\tau)u- (\partial_x+\omega\partial_\xi) (\kappa q+f)+\nu u)=\nu_1 (\partial_x+\omega\partial_\xi)^2u;&
  \label{TxEq2Scl}\\
  &\omega((\partial_t+\omega\partial_\tau)p+(\partial_x+\omega\partial_\xi)(up))=\nu_2 (\partial_x+\omega\partial_\xi)^2p;&
  \label{PrdEq2Scl}\\
  &(\partial_t+\omega\partial_\tau)q-q(1-p-q)=\delta_q (\partial_x+\omega\partial_\xi)^2q.&
  \label{PryEq2Scl}
\end{eqnarray}
 We look for an asymptotic  expansion of the solution to system~(\ref{TxEq2Scl})-(\ref{PryEq2Scl}) that reads as
 \begin{equation}\label{AsmptSrs}
    (u,p,q)=\sum_{k\geqslant 0}\omega^{-k}(u_k,p_k,q_k)(x,t,\xi,\tau), \omega\to\infty.
\end{equation}
We require all the coefficients of series~(\ref{AsmptSrs}) to be  $2\pi-$periodic in $\xi$ and $\tau$.  Replacing the unknowns of system~(\ref{TxEq2Scl})-(\ref{PryEq2Scl}) by series~(\ref{AsmptSrs})  and collecting the terms of equal order in $\omega$ yield a sequence of equations, which we will be solving step by step.

Collecting the terms of order $\omega^2$ in equation~(\ref{PryEq2Scl}) leads to the equation
\begin{equation}\label{EqQ0}
  \partial_{\xi\xi}q_{0}=0.
\end{equation}
There are  no periodic  solutions  to equation (\ref{EqQ0}) except for those independent of $\xi$.  Hence
\begin{equation}\label{q0=}
q_0={q}_0(x,t,\tau).
\end{equation}
Function ${q}_0$ remains unknown to us, and we have to determine it at the subsequent steps.
 Given the equality (\ref{q0=}),  collecting the terms of order $\omega^2$ in equations (\ref{TxEq2Scl}-\ref{PrdEq2Scl}) leads to the equations
\begin{eqnarray}
& (\partial_\tau-\nu_1\partial_{\xi\xi})u_0=\partial_\xi f; &
\label{EqU0}\\
&(\partial_\tau-\nu_2\partial_{\xi\xi})p_0+\partial_\xi(u_0p_0)= 0;&
\label{EqP0}\\
&(u_0,p_0)=(u_0,p_0)(x,t,\xi,\tau),\ (\xi,\tau)\in \mathbb{T}^2.&
\nonumber
\end{eqnarray}
Note that equation~(\ref{EqU0}) is exactly the first equation in problem~(\ref{Tldu0Eq}). Equation~(\ref{EqU0}) has only one periodic solution vanishing on average in the sense of definition~(\ref{DfnAvr}). We denote this solution as $\tilde{u}_0$. Thus  $u_0=\bar{u}+\tilde{u}_0$, $\bar u=\langle u_0\rangle$, and we have justified the leading term in the  asymptotic approximation for $u$ given by  (\ref{u=}), (\ref{Tldu0Eq}).

We  need the following
\\
\textbf{Lemma.} {\sl Let $w=w(\xi,\tau)$ be a smooth  function on $\mathbb{T}^2$. Consider equation}
\begin{equation}\label{QEq}
\partial_\tau Q+\partial_\xi(wQ-\eps\partial_{\xi}Q)=0\ \text{on}\ \mathbb{T}^2;\ \eps=\const>0.
\end{equation}
{\sl Then there exists a unique $2\pi$-periodic (in $\xi$ and $\tau$)  solution to equation~(\ref{QEq}) satisfying the additional condition}
\begin{equation}\label{QAvr}
    \langle Q\rangle=1.
\end{equation}
We'll prove this assertion at the end of this Appendix.

We continue constructing the asymptotic expansion. By the above lemma, problem~(\ref{PEq}) has a unique solution $P=P(x,t,\xi,\tau)$. Hence every solution to  equation~(\ref{EqP0}) reads as
\begin{equation}\label{p0=}
    p_0=\bar{p}(x,t)P(x,t,\xi,\tau).
\end{equation}
 Thus we have got  the leading term of asymptotic  approximation~(\ref{p=}) for unknown $p$.

Now let us consider the terms of order $\omega$.  It follows from equations~(\ref{PryEq2Scl}) and (\ref{q0=}) that
\begin{equation}\label{EqQ1}
    \delta_q\partial_{\xi\xi}q_1=\partial_\tau q_0.
\end{equation}
 Since function $q_0$ does not depend on $\xi$,  equation (\ref{EqQ1}) has a periodic solution if and only if function $q_0$ does not depend on $\tau$. Hence $q_0=\bar{q}(x,t)$ and we get the leading term of asymptotic approximation~(\ref{q=}) for unknown $q$. Further,  every  solution to equation~(\ref{EqQ1}) reads as
\begin{equation}\label{q1=}
    q_1=q_{1}(x,t,\tau).
\end{equation}
Functions $q_1$ and $\bar{q}$ remains unknown to us, and we have to determine them  at the subsequent steps. Note that  the existence of a periodic solution to equation~(\ref{EqQ1}) justifies the error estimate (i.e. $O-$ term) of  asymptotic approximation~(\ref{q=}).

Given the equality~(\ref{q1=}), collecting the terms   of order $\omega$ in equation~(\ref{TxEq2Scl})-(\ref{PrdEq2Scl}) leads to the following equations
\begin{eqnarray}
&(\partial_\tau u_1-\nu_1\partial_{\xi\xi})u_1=2\nu_1\partial_{x\xi}u_0+\partial_x  (\kappa q_0+ f)-\nu u_0-\partial_t u_0,\quad (\xi,\tau)\in \mathbb{T}^2;&
\label{EqU1}\\
 & \partial_\tau p_1+\partial_\xi(u_0p_1)-\nu_2\partial_{\xi\xi}p_1=2\nu_2\partial_{x\xi} p_0-\partial_t p_0-\partial_x(u_0p_0)-\partial_\xi(u_1p_0),\quad (\xi,\tau)\in \mathbb{T}^2.&
\label{EqP1}
\end{eqnarray}
Averaging equations~(\ref{EqU1}) and (\ref{EqP1}) while keeping in mind equality~(\ref{p0=}) brings us at equations~(\ref{EqUSlw0})-(\ref{EqPSlw0}).  Similarly, proceeding with equation (\ref{PryEq2Scl}), we get equation (\ref{EqQSlw0}), that completes the homogenized system consisting of equations (\ref{EqUSlw0})-(\ref{EqQSlw0}). Resolving the homogenized system implies the existence  of periodic solutions to equations~(\ref{EqU1}) and (\ref{EqP1}), which, in turn,  justifies $O-$terms of the asymptotic approximations~(\ref{u=}) and (\ref{p=})).

Now we  pass to proving  the  lemma.
Let $\mathrm{H}$ be the space of the Fourier series in $\xi,\tau$ with  square-summable  coefficients and let $\mathcal{L}:\mathrm{H}\to\mathrm{H}$ be operator defined by the left-hand side of equation~(\ref{QEq}). We have to prove that
\begin{equation}\label{LmmExst}
    \dim\,\mathrm{Ker}\,\mathcal{L}=1, \ \langle \chi\rangle\neq 0\ \forall \ \chi\in \mathrm{Ker}\,\mathcal{L}\setminus\{0\}.
\end{equation}
Let $\mathcal{L}^*$ denote the operator adjoint  to $\mathcal{L}$ and let $\mathcal{J}:\mathrm{H}\to\mathrm{H}$ be the action of inversion $(\xi,\tau)\mapsto(-\xi,-\tau)$. Define
$$
\breve{\mathcal{L}}^*=\mathcal{{J}}\mathcal{L}^*\mathcal{{J}}
$$
Then
$$
\breve{\mathcal{L}}^*:\varphi\mapsto (\partial_\tau-\eps\partial_{\xi\xi})\varphi+w\partial_\xi\varphi.
$$
Notice that PDE
$$
(\partial_\tau-\eps\partial_{\xi\xi})\varphi+w\partial_\xi\varphi=0
$$
obeys the strong maximum and minimum principles~(see, e.g. \cite{Nrnbrg} or \cite{Lnds}). Hence
$$
\mathrm{Ker}\,\breve{\mathcal{L}}^*=\{\varphi\equiv\const\}=\mathrm{Ker}\,\mathcal{L}^*.
$$
Applying the unilateral strong maximum/minimum principles to PDE
$$
(\partial_\tau-\eps\partial_{\xi\xi})\varphi+w\partial_\xi\varphi=1
$$
shows that neither equation $\breve{\mathcal{L}}^*\breve{\psi}=1$ nor equation
$
{\mathcal{L}}^*\psi=1
$
has a solution belonging to $\mathrm{H}$.  Consequently,  the resolvent   $(\mathcal{L}^*-\lambda \mathcal{I})^{-1}$, $\lambda\in \mathbb{C}$,  has a simple pole at  the origin. Since this resolvent is compact,  the pair of operators $\mathcal{L}^*$ and $\mathcal{L}$ obeys the Fredholm theorems. Hence $\dim\,\mathrm{Ker}\,\mathcal{L}=1$. Furthermore, conjecturing that $\langle \chi\rangle=0$ for some $\chi\in \mathrm{Ker}\,\mathcal{L}\setminus\{0\}$  would imply the existence of solution to equation $\mathcal{L}^*\psi=\const\neq 0$ but this contradicts to what we have proved above. This completes the proof.
 \begin{Remark}\label{RmOnPrctr}  Let $\Pi^*$ denote spectral projector onto $\mathrm{Ker}\,\mathcal{L}^*$. Let $\Pi^*(\sigma)$ denote the family of such projectors induced by the conjugated to equation~(\ref{SgmToP}).  Then the action of mapping (\ref{PMpng}) is identical to $\eta\mapsto \langle\Pi^*(\sigma)\tilde{u}_0\rangle$. Hence the perturbation theory for the linear operators implies that this mapping is differentiable and even analytic in the vicinity of origin. Therefore, the drift operator is analytic too by the definition of it by equality (\ref{Drft}), and the derivatives of it  can be evaluated using equalities~(\ref{DffV})-(\ref{P1onXi}).
 \end{Remark}
\section{Appendix II. Linear stability. }
\setcounter{equation}{0}
\noindent
In this appendix, we put the details of the linear stability analysis of the equilibria and quasi-equilibria.
\subsection{Equlibria}
\noindent
\label{ApndxStbElbr}
Let us choose an equilibrium out of family (\ref{Eqlbr}) by specifying the value of the family parameter denoted as $p_e$.
The eigenvalues corresponding to eigenmode~(\ref{EgnMds}) having a specific  wavenumber $\alpha$ are solutions to the following algebraic equation
\begin{eqnarray}
&\lambda^3+(D_1+D_2+D_3)\lambda^2+(D_2D_3+D_1D_3+D_1D_2)\lambda +D_1D_2D_3+\beta\kappa p_e q_e=0,
& \label{ChrEq} \\
&D_1 = \nu+\beta\delta_u;\ D_2=\beta\delta_p;\ D_3=q_e+\beta\delta_q;\ \beta=\alpha^2,\ p_e+q_e=1&
\nonumber
\end{eqnarray}
 By restrictions~(\ref{Restr}), all the coefficients of the polynomial on the left-hand side of equation~(\ref{ChrEq}) are strictly positive. Consequently, the roots of this polynomial are neither positive nor zero. Hence neither unstable nor neutral eigenmode corresponds to a real eigenvalue.

It follows from the Routh-Hurwitz theorem that the necessary and sufficient condition for belonging all the roots of polynomial (\ref{ChrEq}) to the open  left half-plane of the complex plane reads as
$$
(D_1+D_2+D_3)(D_2D_3+D_1D_3+D_1D_2)>D_1D_2D_3+\beta\kappa p_e q_e.
$$
This inequality allows a more compact form, namely:
$$
(D_1+D_2)(D_1+D_3)(D_2+D_3)>\beta\kappa p_e q_e.
$$
Since the degree of  polynomial (\ref{ChrEq}) is 3,   replacing the last inequality by equality is necessary and sufficient for belonging the root of the polynomial (\ref{ChrEq}) to the imaginary axis, and this root cannot be zero. Given  this observation, we conclude that the critical magnitude of predators' motility discussed in section~\ref{SSecDstbEqlbr} has the following form
\begin{equation}\label{KppaCr}
   \kappa_c=\frac{(D_1+D_2)(D_1+D_3)(D_2+D_3)}{\beta p_e q_e}.
\end{equation}
In the case of  $\delta_u=\delta_p=0$,  which is of particular interest for the  considerations of section~\ref{SSecDstbQsEqlbr}, expression~(\ref{KppaCr})   simplifies to
\begin{equation}\label{Kppa0Cr}
    \kappa_c^0=\kappa_c|
    _{\delta_p=\delta_u=0}={\frac {{  \nu}\, \left( {  \delta_q}\,\beta+{  \nu}+q_e \right)  \left( {
  \delta_q}\,\beta+q_e \right) }{q_e\, p_e \beta}}.
\end{equation}
The direct inspection using the expression ~(\ref{KppaCr}) or (\ref{Kppa0Cr})  shows the positiveness of the threshold value of motility denoted as  $\kappa_*(\delta)$ in section~\ref{SSecDstbEqlbr}. For instance,
$$
\left.\kappa_*(\delta)\right|_{\delta=(1,1,0,0)}=27/2.
$$
\subsection{Quasi-equilibria}
\label{AppndxStbltQsEqlbr}
\noindent
Given the external signal having the form of travelling wave~(\ref{TrWvUnMdltd}), we analyze the linear stability of a homogeneous quasi-equilibrium.  Then the small perturbations of it obey system~(\ref{TxEqLnrSlw})-(\ref{PrdEqLnrSlw}), all the coefficients of which are constant.  Acting  the   same way as in the case of equilibria, we arrive at the characteristic polynomial, that reads as
$$
{\lambda}^{3}+ \left( iU+\nu+q_e+\delta_q\beta \right) {\lambda}^{2}+ \left(  \left( i
U+q_e+\delta_q\beta \right) \nu+iU (q_e+\delta_q\beta)\,  \right) \lambda+i\nu U (q_e+\delta_q\beta)\, +\beta p_eq_e\bar{\kappa}.
$$
Here $\beta=\alpha^2$, $\alpha$ stands for the eigenmode wavenumber, $U=\alpha v_e$,  $v_e$ stands for the drift velocity determined by equality (\ref{TrWvDrftVlct}), and $\bar\kappa=(1+\mathcal{V}^\prime(f))\kappa$ is the effective motility determined by equality (\ref{TrWvMtltFctr}). For the sake of convenience, we write this polynomial in more compact form, namely,
$$
{\lambda}^{3}+ \left( iU+d_1\right) {\lambda}^{2}+\left( d_2+iU d_1\,  \right) \lambda+i Ud_2\, +r 
$$
where
$$
d_1=\nu+q_e+\delta_q\beta,\ d_2=\nu(q_e+\delta_q\beta),\ r=\beta p_eq_e\bar{\kappa}.
$$
Note that $r>0$, $d_1>0$ and  $d_2>0$ for every admissible set of the problem parameters\footnote{The admissibility of the problem parameters presumes the positiveness of the values of $\kappa$ and, therefore,  $ \bar\kappa$.}.

We employ the complex-valued  version of Hurwitz's theorem  to count  the  roots belonging to the right complex semi-plane, and  it brings us at  the chain of  the Hankel's matrix minors, that reads as
$$
1,\ d_1,\
d_1\left( d_2d_1-r\right),\ r\left((r-d_1d_2)^2 -{U}^{2}{{  d_1}}^{2}{  d_2}\right).
$$
  For this chain, there are four generic distributions of  signs of the minors, namely
$$
++++,\quad +++-,\quad++--,\quad++-+.
$$
Switching between the diagrams displayed herein corresponds to crossing certain critical submanifolds in the space of the problem parameters.  These are the null sets of functions $\bar{\kappa}-\kappa_c^{(j)}(v_e,p_e,\beta,\delta)$, $j\in\{+,-,0\}$, where $\kappa_c^{(0)}$ is the critical motility defined by equality (\ref{Kppa0Cr}),  and
\begin{equation}\label{KppaCrPM}
    \kappa_c^{(\pm)}= \frac{c_*d_1(c_* \pm |v_e|)}{p_eq_e},\ c_* =\sqrt{{d_2}/{\beta}},
\end{equation}
Note that  $\kappa_c^{-}<\kappa_c^0<\kappa_c^{+}$ for every admissible  set of the problem parameters provided that $v_e\neq 0$. If $v_e=0$ then $\kappa_c^{-}=\kappa_c^0=\kappa_c^{+}$. Therefore, passing the effective motility through the values of $\kappa_c^{-}$, $\kappa_c^0$, and $\kappa_c^{+}$ in the ascending order   causes switching between the above diagrams from left to right provided that the values of all other parameters stay unaltered. Hence the number of roots in the right complex semi-plane takes the values of 0,1 and 2 provided that the value of  $\bar{\kappa}$ belongs to  intervals $(0,\kappa_c^-)$, $(\kappa_c^-,\kappa_c^+)$ and $(\kappa_c^+,\infty)$ correspondingly.  Changing this number is due to changing the sign of the senior minor of Hankel's matrix. Therefore,  passing the effective motility through one of the values of $\kappa_c^{\pm}$ causes crossing the imaginary axis by the root of the characteristic polynomial at some non-zero point. Hence the oscillatory instability occurs at this moment.
Note that passing through the value of  $\kappa_c^0$ does not change the number of unstable roots.

For $\kappa=\kappa^\pm_c$ and $v_e>0$ the neutral eigenmode  is proportional to
$$
\mathrm{e}^{i\sqrt{\beta}(x\pm c_* t)},
$$
and every eigenmode appears together with the complex-conjugated one. For  $v_e<0$, it suffices to substitute $x\pm c_* t$ by $x \mp c_* t$.
\begin{Remark}
 \label{RmOnOmgVsVe}
Inequalities $|v_e|\gtrless c_*$ are equivalent to $\kappa_c^-\lessgtr 0$. Occurring the negative values of $\kappa_c^-$ for a specific wavelength excludes the diagram $++++$. Hence the eigenmode with such a wavelength is unstable for every  quasi-equilibrium.
 \end{Remark}
\section{Appendix III. Estimates of exponential integrals. }
\setcounter{equation}{0}
\noindent
 In this appendix, we estimate the integrals introduced in~(\ref{TrWvGm}) and denoted as $\Gamma_\pm$ when the signal amplitude denoted as $a$ tends to infinity.  For simplicity, we restrict ourselves within the class of  the unmodulated travelling waves defined by equality (\ref{TrWvUnMdltd}). Then $\Gamma_\pm=\Gamma_\pm(z),\ z\gtrless 0$. For definiteness, let's consider   $\Gamma_+=\Gamma_+(z), z>0$. We put this integral into a slightly different form  that is more convenient for applying the Laplace method, namely,
\begin{equation}\label{ExpInt}
    \frac{1}{2\pi}\int\limits_{(0,\infty)\times(0,2\pi)} \mathrm{e}^{-z\sigma}\mathrm{e}^{aS(\eta,\sigma)}d\eta d\sigma, \quad S(\eta,\sigma)=s(\eta)-s(\eta-\sigma),\ z>0
\end{equation}
For simplicity, we consider function $s$ as a given one, and we assume that it is  $2\pi-$periodic, vanishing on average and analytic and that every critical point of it is non-degenerated.  The periodicity allows us to reduce the domain of integration of integral  (\ref{ExpInt}) to cylinder
$$
C=\{(\eta,\sigma) \in \mathbb{S}\times(0,2\pi)\},
$$
and we arrive at estimating the following integral
\begin{equation}\label{ExpInt1}
\int\limits_{C} \mathrm{e}^{-{z}\sigma}\mathrm{e}^{aS(\eta,\sigma)}d\eta d\sigma .
\end{equation}
The local maximizers of $S$ on the  whole $(\eta,\sigma)-$plain are
$$
(\eta,\sigma)\in\mathbb{R}^2:\ s^\prime(\eta)=s^\prime(\eta-\sigma)=0,\quad  s^{\prime\prime}(\eta)s^{\prime\prime}(\eta-\sigma)<0.
$$
The last inequality  means that no maximizers belong to $\pr C$. Then the leading term in Laplace's asymptotics of integral (\ref{ExpInt1}) for $a\to\infty$ reads as
$$
\frac{2\pi \mathrm{e}^{a\mathrm{osc}(s)}}{a}\sum\limits_{(x,y)\in M} \frac{\mathrm{e}^{-{z}(x-y)}}{\sqrt{-s^{\prime\prime}(x)s^{\prime\prime}(y)}}
$$
where
\begin{equation}\label{DfOscAndSetM}
 \mathrm{osc}(s)=\sup\limits_\mathbb{R} s-\inf\limits_\mathbb{R} s,\quad M=  \{(x,y):\ 0<x\le 2\pi, 0<x-y<2\pi, \mathrm{osc}(s)=s(x)-s(y)\}.
\end{equation}
Coming back to the integral (\ref{ExpInt}) brings us at the following estimate
\begin{equation}\label{GnrlLplcEst}
    \Gamma_+(z)\sim \frac{ \mathrm{e}^{a\mathrm{osc}(s)}}{a(1-\mathrm{e}^{-2\pi {z}})}\sum\limits_{(x,y)\in M} \frac{\mathrm{e}^{-{z}(x-y)}}{\sqrt{-s^{\prime\prime}(x)s^{\prime\prime}(y)}}, \quad a\to \infty.
\end{equation}
\begin{Exmpl}
\label{ExmplGnrlLplcEstSn}
Let $s(\eta)=\sin\eta$. Then $\mathrm{osc}(s)=2$, $M=\{(\pi/2, -\pi/2)\}$, and estimate~(\ref{GnrlLplcEst}) reads as
$$
\Gamma_+(z)\sim\frac{ \mathrm{e}^{2a}\mathrm{e}^{-\pi{z}}}{a(1-\mathrm{e}^{-2\pi {z}})}, \quad a\to \infty.
$$
\end{Exmpl}
\begin{Remark}
\label{RmrkGnrlLplcEstCTo0}
Asymptotics (\ref{GnrlLplcEst}) delivers an estimate for the residual drift velocity that we have been defining by  equality (\ref{TrWvRsdlDrft}) involving  the value of $\Gamma_+$ for $z=c/\nu_2$.  This estimate matches the limit identity (\ref{DrftVlctCTo0Infty}).  In particular, using the estimate of example~\ref{ExmplGnrlLplcEstSn} for $z\to+0$  bring us at the leading term in the asymptotics of function $I^2_0(a)$, $a\to\infty$, in consistence with what we have learned  from the example~\ref{Exmplsn}.  Note finally that differentiating the asymptotics of $\Gamma_+$  gives us the effective motility factor defined by equality (\ref{TrWvMtltFctr}).
\end{Remark}
\end{document}